\documentclass[12pt]{article}
\usepackage{a4wide}
\usepackage{latexsym}
\usepackage{amsmath}
\usepackage{amsfonts}
\usepackage{amscd}
\usepackage{cite}
\usepackage{graphicx}
\usepackage{float}

\usepackage{pslatex}
\usepackage[latin1]{inputenc}
\usepackage[T1]{fontenc}

\def\bq{\begin{eqnarray}}
\def\eq{\end{eqnarray}}
\def\l{\langle}
\def\r{\rangle}

\begin{document}

\thispagestyle{empty}

\begin{flushright}
  MZ-TH/09-35
\end{flushright}

\vspace{1.5cm}

\begin{center}
  {\Large\bf Moments of event shapes in electron-positron annihilation at NNLO\\
  }
  \vspace{1cm}
  {\large Stefan Weinzierl\\
\vspace{2mm}
      {\small \em Institut f{\"u}r Physik, Universit{\"a}t Mainz,}\\
      {\small \em D - 55099 Mainz, Germany}\\
  } 
\end{center}

\vspace{2cm}

\begin{abstract}\noindent
  {
This article gives the perturbative NNLO results for the moments of the
most commonly used event shape variables associated to three-jet events
in electron-positron annihilation: Thrust, heavy jet mass, wide jet broadening,
total jet broadening, C parameter and the Durham three-to-two jet transition variable.
   }
\end{abstract}

\vspace*{\fill}

\newpage

\section{Introduction}
\label{sec:intro}

The electron-positron annihilation 
experiments at the colliders LEP (CERN), SLC (SLAC) and PETRA (DESY) have collected a wealth of data 
with hadronic final state over a wide range of energies.
Of particular interest are three-jet events, which can be used to extract the value of the strong coupling.
Three-jet events are well suited for this task because the leading term in a perturbative calculation 
of three-jet observables is already proportional to the strong coupling.
In comparing experiments to theory it is important to restrict oneself to infra-red safe observables.
For three-jet events in electron-positron annihilation there is a well established set of infra-red safe observables, 
which is widely used. These are the event shape variables consisting of thrust, heavy jet mass, wide jet broadening,
total jet broadening, the $C$-parameter and the Durham three-to-two jet transition variable.
All these observables can be calculated in perturbation theory. 
Next-to-next-to-leading order (NNLO) results for the distributions of these observables have been 
presented in \cite{GehrmannDeRidder:2007hr,GehrmannDeRidder:2007bj,GehrmannDeRidder:2008ug,Weinzierl:2009ms}.
Apart from the distributions also the moments of these observables are of interest to the experimentalists.
For an observable ${\cal O}$ which takes values between $0$ and $1$ the $n$-th moment is defined by
\bq 
 \left\langle {\cal O}^n \right\rangle
 & = &
 \frac{1}{\sigma_{tot}} \int\limits_0^1 {\cal O}^n \frac{d\sigma}{d{\cal O}} d{\cal O}.
\eq
All observables are chosen such that $O \rightarrow 0$ corresponds to the two-jet region.
On the other hand, large values of $O$ correspond to the multi-jet region.
For all $n$ the $n$-th moment receives contributions from both regions, but
the two regions are weighted differently for different $n$:
For higher $n$ more weight is given to the multi-jet region.

In this article I present the QCD NNLO results for the moments of the event shape observables.
The present calculation is based on the numerical Monte Carlo program Mercutio2 
\cite{Weinzierl:2009nz,Weinzierl:2008iv,Weinzierl:1999yf}.
Independent results for the moments have been published in \cite{GehrmannDeRidder:2009dp}.

The results of this paper rely heavily on research carried out in the past years related to differential NNLO calculations:
Integration techniques for two-loop amplitudes \cite{Gehrmann:1999as,Gehrmann:2000zt,Gehrmann:2001ck,Moch:2001zr,Weinzierl:2002hv,Bierenbaum:2003ud,Weinzierl:2004bn,Moch:2005uc},
the calculation of the relevant tree-, one- and two-loop-amplitudes \cite{Berends:1989yn,Hagiwara:1989pp,Falck:1989uz,Schuler:1987ej,Korner:1990sj,Giele:1992vf,Bern:1997ka,Bern:1997sc,Campbell:1997tv,Glover:1997eh,Garland:2001tf,Garland:2002ak,Moch:2002hm},
routines for the numerical evaluation of polylogarithms \cite{Gehrmann:2001pz,Gehrmann:2001jv,Vollinga:2004sn},
methods to handle infrared singularities \cite{Catani:1997vz,Phaf:2001gc,Catani:2002hc,Kosower:2002su,Kosower:2003cz,Weinzierl:2003fx,Weinzierl:2003ra,Kilgore:2004ty,Frixione:2004is,Gehrmann-DeRidder:2003bm,Gehrmann-DeRidder:2004tv,Gehrmann-DeRidder:2005hi,Gehrmann-DeRidder:2005aw,Gehrmann-DeRidder:2005cm,GehrmannDeRidder:2007jk,Somogyi:2005xz,Somogyi:2006da,Somogyi:2006db,Catani:2007vq,Somogyi:2008fc,Somogyi:2009ri,Aglietti:2008fe}
and experience from the NNLO calculations of
$e^+ e^- \rightarrow \mbox{2 jets}$ and other processes
\cite{Anastasiou:2004qd,Gehrmann-DeRidder:2004tv,Weinzierl:2006ij,Weinzierl:2006yt,Anastasiou:2002yz,Anastasiou:2003yy,Anastasiou:2003ds,Anastasiou:2004xq,Anastasiou:2005qj,Anastasiou:2007mz,Melnikov:2006di,Anastasiou:2005pn,Catani:2001ic,Catani:2001cr,Grazzini:2008tf,Harlander:2001is,Harlander:2002wh,Harlander:2003ai,Ravindran:2003um,Ravindran:2004mb}.

This article reports the pure QCD perturbative results for the moments of the event shapes.
Not included are soft-gluon resummations \cite{Catani:1992ua,Catani:1998sf,Becher:2008cf,Gehrmann:2008kh}
nor power corrections \cite{Dokshitzer:1997ew,Davison:2008vx}.
Perturbative electro-weak corrections to three-jet observables have been reported recently 
in \cite{CarloniCalame:2008qn,Denner:2009gx}.

From a technical point of view the calculation of the $n=1$ moment is particular challenging. 
The first moment receives sizable contributions from the close-to-two-jet region.
The calculation of the NNLO correction in the close-to-two-jet region involves the integration over 
three unresolved partons -- one additional unresolved parton more as compared to a ``standard'' NNLO calculation.
The integration is done numerically by Monte Carlo methods and highly non-trivial.

This paper is organised as follows: In the next section the set of event shape variables are introduced.
In section~\ref{sec:perturbative} the perturbative calculation is described.
The numerical results for the moments are given in section~\ref{sec:num}.
Finally, section~\ref{sec:conclusions} contains the conclusions.

\section{Definition of the observables}
\label{sec:def}

In this section I briefly recall the definitions of the relevant event shape observables.
The event shape variable thrust \cite{Brandt:1964sa,Farhi:1977sg}
is defined by
\bq
 T & = & 
 \frac{\max\limits_{\vec{n}} \sum\limits_j \left| \vec{p}_j \cdot \vec{n} \right|}{\sum\limits_j \left| \vec{p}_j \right|},
\eq
where $\vec{p}_j$ denotes the three-momentum of particle $j$ and the sum runs over all particles in the 
final state. The thrust variable maximises the total longitudinal momentum along the unit vector $\vec{n}$.
The value of $\vec{n}$, for which the maximum is attained is called the thrust axis
and denoted by $\vec{n}_T$. 
The value of thrust ranges between $1/2$ and $1$, where $T=1$ corresponds to an ideal collinear two-jet event
and $T=1/2$ corresponds to a perfectly spherical event.
Usually one considers instead of thrust $T$ the variable $(1-T)$, such that
the two-jet region corresponds to $(1-T) \rightarrow 0$.

The plane orthogonal to the thrust axis divides the space into two hemispheres $H_1$ and $H_2$. These are used
to define the following event shape variables: 
The hemisphere masses \cite{Clavelli:1981yh} are defined by
\bq
 M_i^2 & = & \left( \sum\limits_{j\in H_i} p_j \right)^2, \;\;\;i=1,2,
\eq
where $p_j$ denotes the four-momentum of particle $j$.
The heavy hemisphere mass $M_H$ is then defined by
\bq
 M_H^2 = \max\left(M_1^2,M_2^2\right).
\eq
It is convenient to introduce the dimensionless quantity
\bq
 \rho & = & \frac{M_H^2}{Q^2},
\eq
where $Q$ is the centre-of-mass energy.
In leading order the distribution of the heavy jet mass $\rho$ is identical to the distribution of $(1-T)$.
Experimentalists also use the square root of $\rho$ as observable:
\bq
 \sqrt{\rho} & = & \frac{M_H}{Q}.
\eq
In this paper we provide in addition to the results for the moments of $\rho$ also the results for the moments
of $\sqrt{\rho}$.
The odd moments of $\sqrt{\rho}$ give new information, while the even moments of $\sqrt{\rho}$ are related to the
moments of $\rho$ as follows:
The $2n$-th moment of $\sqrt{\rho}$ is identical to the $n$-th moment of $\rho$.

The hemisphere broadenings \cite{Rakow:1981qn,Catani:1992jc} are defined by
\bq
 B_i & = & 
 \frac{\sum\limits_{j\in H_i} \left| \vec{p}_j \times \vec{n}_T \right|}
      {2 \sum\limits_k \left| \vec{p}_k \right|},
 \;\;\;i=1,2,
\eq
where the sum over $j$ runs over all particles in one of the hemispheres, whereas the sum over $k$ is over all
particles in the final state.
The wide jet broadening $B_W$ and the total jet broadening $B_T$ are defined
by
\bq
 B_W = \max\left(B_1,B_2\right), 
 \;\;\;
 B_T = B_1 + B_2.
\eq

The $C$-parameter \cite{Parisi:1978eg,Donoghue:1979vi} is obtained from the
linearised momentum tensor
\bq
\theta^{ij} & = & 
 \frac{1}{\sum\limits_l \left|\vec{p}_l\right|} \sum\limits_k \frac{p_k^i p_k^j}{\left|\vec{p}_k\right|},
 \;\;\; i,j=1,2,3,
\eq
where the sum runs over all final state particles and $p_k^i$ is the $i$-th component of the three-momentum
$\vec{p}_k$ of particle $k$ in the c.m. system. 
The tensor $\theta$ is normalised to have unit trace.
In terms of the eigenvalues of the $\theta$ tensor,
$\lambda_1$, $\lambda_2$, $\lambda_3$, with
$\lambda_1 + \lambda_2 + \lambda_3 = 1$, one defines
\bq
C & = & 3 \left( \lambda_1 \lambda_2 + \lambda_2 \lambda_3 + \lambda_3 \lambda_1 \right).
\eq
The $C$-parameter exhibits in perturbation theory a singularity at the 
three-parton boundary $C=3/4$ \cite{Catani:1997xc}.

For the three-to-two jet transition variable $y_{23}$ one first defines jets according to the 
Durham jet algorithm \cite{Stirling:1991ds}:
The clustering procedure of the jet algorithm 
is defined through the following steps:
\begin{enumerate}
\item Define a resolution parameter $y_{cut}$
\item For every pair $(p_k, p_l)$ of final-state particles compute the corresponding
resolution variable $y_{kl}$.
\item If $y_{ij}$ is the smallest value of $y_{kl}$ computed above and
$y_{ij} < y_{cut}$ then combine $(p_i, p_j)$ into a single jet ('pseudo-particle')
with momentum $p_{ij}$ according to a recombination prescription.
\item Repeat until all pairs of objects (particles and/or pseudo-particles)
have $y_{kl} > y_{cut}$.
\end{enumerate}
For the Durham algorithm the resolution variable is given by
\bq
y_{ij} = \frac{2 \min(E_i^2, E_j^2) \left( 1 - \cos \theta_{ij} \right)}{Q^2},
\eq
while the recombination prescription is given by
\bq
E_{ij} = E_i + E_j,
 & &
\vec{p}_{ij} = \vec{p}_i + \vec{p}_j.
\eq
Here, $E_i$ and $E_j$ are the energies of particles $i$ and $j$, and $\theta_{ij}$ is the angle
between the three-momenta $\vec{p}_i$ and $\vec{p}_j$.
$Q$ is the centre-of-mass energy.
The jet transition variable $y_{23}$ is the value of the jet resolution parameter
$y_{cut}$, for which the event changes from a three-jet to a two-jet configuration.

\section{Perturbative expansion}
\label{sec:perturbative}

The perturbative expansion of a differential distribution weighted by the $n$-th power of the observable $\cal O$ 
can be written 
for any infrared-safe observable for the process
$e^+ e^- \rightarrow \mbox{3 jets}$
up to NNLO as
\bq
\frac{{\cal O}^n}{\sigma_{tot}(\mu)} \frac{d\sigma(\mu)}{d{\cal O}} & = & 
 \frac{\alpha_s(\mu)}{2\pi} \frac{{\cal O}^n d\bar{A}_{\cal O}(\mu)}{d{\cal O}}
 +
 \left( \frac{\alpha_s(\mu)}{2\pi} \right)^2 \frac{{\cal O}^n d\bar{B}_{\cal O}(\mu)}{d{\cal O}}
 +
 \left( \frac{\alpha_s(\mu)}{2\pi} \right)^3 \frac{{\cal O}^n d\bar{C}_{\cal O}(\mu)}{d{\cal O}}.
\eq
$\bar{A}_{\cal O}$ gives the LO result, $\bar{B}_{\cal O}$ the NLO correction and $\bar{C}_{\cal O}$ the NNLO correction.
$\sigma_{tot}$ denotes the total hadronic cross section calculated up to the relevant order.
The arbitrary renormalisation scale is denoted by $\mu$.
All observables are chosen such that they take values between $0$ and $1$.
The $n$-th moment is given by
\bq 
 \left\langle {\cal O}^n \right\rangle
 & = &
 \frac{1}{\sigma_{tot}} \int\limits_0^1 {\cal O}^n \frac{d\sigma}{d{\cal O}} d{\cal O}.
\eq
In practise the numerical program computes the distribution
\bq
\frac{{\cal O}^n}{\sigma_{0}(\mu)} \frac{d\sigma(\mu)}{d{\cal O}} & = & 
 \frac{\alpha_s(\mu)}{2\pi} \frac{{\cal O}^n dA_{\cal O}(\mu)}{d{\cal O}}
 +
 \left( \frac{\alpha_s(\mu)}{2\pi} \right)^2 \frac{{\cal O}^n dB_{\cal O}(\mu)}{d{\cal O}}
 +
 \left( \frac{\alpha_s(\mu)}{2\pi} \right)^3 \frac{{\cal O}^n dC_{\cal O}(\mu)}{d{\cal O}},
\eq
normalised to $\sigma_0$, which is the LO cross section for $e^+ e^- \rightarrow \mbox{hadrons}$,
instead of the normalisation to $\sigma_{tot}$.
There is a simple relation between the two distributions:
The functions $A_{\cal O}$, $B_{\cal O}$ and $C_{\cal O}$ are related to the functions
$\bar{A}_{\cal O}$, $\bar{B}_{\cal O}$ and $\bar{C}_{\cal O}$ by
\bq
\label{rel_bar}
 \bar{A}_{\cal O} & = & A_{\cal O},
 \nonumber \\
 \bar{B}_{\cal O} & = & B_{\cal O} - A_{tot} A_{\cal O} ,
 \nonumber \\
 \bar{C}_{\cal O} & = & C_{\cal O} - A_{tot} B_{\cal O} - \left( B_{tot} - A_{tot}^2 \right) A_{\cal O} ,
\eq
where
\bq
A_{tot} & = & \frac{3(N_c^2-1)}{4 N_c},
 \nonumber \\
 B_{tot} & = &   
 \frac{N_c^2-1}{8 N_c} \left[ \left( \frac{243}{4} - 44 \zeta_3 \right) N_c + \frac{3}{4 N_c} 
                             + \left( 8 \zeta_3 - 11 \right) N_f \right]. 
\eq
$N_c$ denotes the number of colours and $N_f$ the number of light quark flavours.
$A_{tot}$ and $B_{tot}$ are obtained from the perturbative expansion of $\sigma_{tot}$ \cite{Dine:1979qh,Chetyrkin:1979bj,Celmaster:1979xr}:
\bq
 \sigma_{tot} & = & 
 \sigma_0 \left( 1 + \frac{\alpha_s}{2\pi} A_{tot}
                   + \left( \frac{\alpha_s}{2\pi} \right)^2 B_{tot}
                   + {\cal O}(\alpha_s^3) \right).
\eq
The perturbative calculation of the inclusive hadronic cross section $\l \sigma \r^{(tot)}$ is actually known
to ${\cal O}(\alpha_s^3)$ \cite{Gorishnii:1991vf,Surguladze:1990tg}, although we need here only the coefficients up to
${\cal O}(\alpha_s^2)$. 

It is sufficient to calculate the functions $\bar{A}_{\cal O}$, $\bar{B}_{\cal O}$ and $\bar{C}_{\cal O}$
for a fixed renormalisation scale $\mu_0$, which can be taken conveniently to be equal to the
centre-of-mass energy: $\mu_0=Q$.
The scale variation can be restored from the renormalisation group equation
\bq
\label{RGE_alpha_s}
 \mu^2 \frac{d}{d\mu^2} \left( \frac{\alpha_S}{2\pi} \right)
 & = & 
 - \frac{1}{2} \beta_0 \left( \frac{\alpha_S}{2\pi} \right)^2
 - \frac{1}{4} \beta_1 \left( \frac{\alpha_S}{2\pi} \right)^3
 - \frac{1}{8} \beta_2 \left( \frac{\alpha_S}{2\pi} \right)^4
 + {\cal O}(\alpha_s^5),
 \\
 \beta_0 & = & \frac{11}{3} C_A - \frac{4}{3} T_R N_f,
 \nonumber \\
 \beta_1 & = & \frac{34}{3} C_A^2 - 4 \left( \frac{5}{3} C_A + C_F \right) T_R N_f,
 \nonumber \\
 \beta_2 & = & \frac{2857}{54} C_A^3 - \left( \frac{1415}{27} C_A^2 + \frac{205}{9} C_A C_F - 2 C_F^2 \right) T_R N_f
             + \left( \frac{158}{27} C_A + \frac{44}{9} C_F \right) T_R^2 N_f^2.
 \nonumber
\eq
The colour factors are defined as usual by
\bq
 C_A = N_c,
 \;\;\;
 C_F = \frac{N_c^2-1}{2 N_c},
 \;\;\;
 T_R = \frac{1}{2}.
\eq
The values of the functions $\bar{A}_{\cal O}$, $\bar{B}_{\cal O}$ and $\bar{C}_{\cal O}$
at a scale $\mu$ are then obtained from the ones at the scale $\mu_0$ by
\bq
\label{scale_depend}
 \bar{A}_{\cal O}(\mu) & = & \bar{A}_{\cal O}(\mu_0),
 \nonumber \\
 \bar{B}_{\cal O}(\mu) & = & \bar{B}_{\cal O}(\mu_0) + \frac{1}{2} \beta_0 \ln\left(\frac{\mu^2}{\mu_0^2}\right) \bar{A}_{\cal O}(\mu_0),
 \nonumber \\
 \bar{C}_{\cal O}(\mu) & = & \bar{C}_{\cal O}(\mu_0) + \beta_0 \ln\left(\frac{\mu^2}{\mu_0^2}\right) \bar{B}_{\cal O}(\mu_0)
                           + \frac{1}{4} \left[ \beta_1 + \beta_0^2 \ln\left(\frac{\mu^2}{\mu_0^2}\right) \right] \ln\left(\frac{\mu^2}{\mu_0^2}\right) \bar{A}_{\cal O}(\mu_0).
\eq
Finally, an approximate solution of eq.~(\ref{RGE_alpha_s}) for $\alpha_s$ is given by
\bq
 \frac{\alpha_s(\mu)}{2\pi}
 & = &
 \frac{2}{\beta_0 L}
 \left\{
         1 - \frac{\beta_1}{\beta_0^2} \frac{\ln L}{L}
         + \frac{\beta_1^2}{\beta_0^4 L^2}
           \left[ \left( \frac{1}{2} - \ln L \right)^2
                  + \frac{\beta_0 \beta_2}{\beta_1^2}
                  - \frac{5}{4}
           \right]
 \right\},
\eq
where $L=\ln(\mu^2/\Lambda^2)$.

The function $C_{\cal O}$ can be decomposed into six colour pieces
\bq
 C_{\cal O} & = &
 \frac{\left( N_c^2-1 \right)}{8 N_c} 
 \left[ 
        N_c^2 C_{\cal O}^{lc}
      + C_{\cal O}^{sc}
      + \frac{1}{N_c^{2}} C_{\cal O}^{ssc}
      + N_f N_c C_{\cal O}^{nf}
      + \frac{N_f}{N_c} C_{\cal O}^{nfsc}
      + N_f^2 C_{\cal O}^{nfnf}
 \right].
\eq
In addition, there are singlet contributions, which arise from interference terms of amplitudes, where
the electro-weak boson couples to two different fermion lines.
These singlet contributions are expected to be numerically 
small \cite{Dixon:1997th,vanderBij:1988ac,Garland:2002ak}
and neglected in the present calculation.
We define
\bq
\label{def_colour_factors}
 & &
  \left. C_{\cal O} \right|_{lc} = \frac{\left( N_c^2-1 \right)}{8 N_c} N_c^2 C_{\cal O}^{lc},
 \;\;\;
  \left. C_{\cal O} \right|_{sc} = \frac{\left( N_c^2-1 \right)}{8 N_c} C_{\cal O}^{sc},
 \;\;\;
  \left. C_{\cal O} \right|_{ssc} = \frac{\left( N_c^2-1 \right)}{8 N_c} \frac{1}{N_c^{2}} C_{\cal O}^{ssc},
 \\
 & &
  \left. C_{\cal O} \right|_{nf} = \frac{\left( N_c^2-1 \right)}{8 N_c} N_f N_c C_{\cal O}^{nf},
 \;\;\;
  \left. C_{\cal O} \right|_{nfsc} = \frac{\left( N_c^2-1 \right)}{8 N_c} \frac{N_f}{N_c} C_{\cal O}^{nfsc},
 \;\;\;
  \left. C_{\cal O} \right|_{nfnf} = \frac{\left( N_c^2-1 \right)}{8 N_c} N_f^2 C_{\cal O}^{nfnf},
 \nonumber 
\eq
e.g. the function $\left. C_{\cal O} \right|_{lc}$ includes the colour factors.

The functions $A_{\cal O}$, $B_{\cal O}$ and $C_{\cal O}$ are calculated for a fixed renormalisation scale equal
to the centre-of-mass energy: $\mu_0=Q$.
They depend only on the value of the observable ${\cal O}$. 
Since only QCD corrections with non-singlet quark couplings are taken into account and 
singlet contributions to $C_{\cal O}$ are neglected, the functions 
$A_{\cal O}$, $B_{\cal O}$ and $C_{\cal O}$ do not depend on electro-weak couplings.

\section{Numerical results}
\label{sec:num}

In this section I present the results for the perturbative coefficients
$A_{{\cal O},n}$, $B_{{\cal O},n}$ and $C_{{\cal O},n}$.
These are related by eq.~(\ref{rel_bar}) to the coefficients 
$\bar{A}_{{\cal O},n}$, $\bar{B}_{{\cal O},n}$ and $\bar{C}_{{\cal O},n}$.
The perturbative expansion of the moments in terms of the latter is given by
\bq
\left\langle {\cal O}^n \right\rangle & = & 
 \frac{\alpha_s}{2\pi} \bar{A}_{{\cal O},n}
 +
 \left( \frac{\alpha_s}{2\pi} \right)^2 \bar{B}_{{\cal O},n}
 +
 \left( \frac{\alpha_s}{2\pi} \right)^3 \bar{C}_{{\cal O},n}
 + {\cal O}\left(\alpha_s^4\right).
\eq
The coefficients $A_{{\cal O},n}$, $B_{{\cal O},n}$ and $C_{{\cal O},n}$ are calculated at the renormalisation scale equal to the 
centre-of-mass energy $\mu_0=Q$.
The scale dependence is given by eq.~(\ref{scale_depend}).

All results have been obtained by numerical Monte-Carlo integration.
The Monte-Carlo integration introduces a statistical error.
For the infrared singularities a hybrid method between subtraction and slicing has been used.
The dimensionless slicing parameter is denoted by
\bq
 \eta & = & \frac{s_{min}}{Q^2}.
\eq
The slicing procedure introduces in addition to the statistical error from the Monte Carlo integration a systematic error.
The size of this error can be estimated by varying the slicing parameter $\eta$.
However, lowering the slicing parameter will increase the statistical error.
A practical criteria is to require that the variation due to the slicing parameter is smaller than
the statistical error, with the possible exception for the boundaries of the distributions.
Imposing this criteria $\eta=10^{-5}$ turns out to be a good compromise between accuracy and efficiency
in the region away from the two-jet region.
However the first moment of the distributions receives non-negligible
contributions from the close-to-two-jet region.
In this region a more careful analysis has to be performed.
The moments are calculated as follows:
We introduce a second parameter $\kappa=10^{-3}$ and split the integral into
\bq
\label{split_up}
 \frac{1}{\sigma_{tot}} \int\limits_0^1 {\cal O}^n \frac{d\sigma}{d{\cal O}} d{\cal O}
 & = & 
 \frac{1}{\sigma_{tot}} \int\limits_0^{\kappa} {\cal O}^n \frac{d\sigma}{d{\cal O}} d{\cal O}
+
 \frac{1}{\sigma_{tot}} \int\limits_{\kappa}^1 {\cal O}^n \frac{d\sigma}{d{\cal O}} d{\cal O}.
\eq
In the region $[\kappa,1]$ the integral is calculated with the slicing parameter $\eta=10^{-5}$, whereas
in the region $[0,\kappa]$ the slicing parameter $\eta=10^{-9}$ is used for the computation of the integral.
In order to estimate the systematical error due to the slicing procedure the moments are recalculated
in the region $[0,\kappa]$ with the slicing parameters $\eta=10^{-7}$ and $\eta=10^{-5}$.
Therefore we obtain three results $I_5$, $I_7$ and $I_9$ for the integral over $[0,\kappa]$,
corresponding to the values $\eta=10^{-5}$, $\eta=10^{-7}$ and $\eta=10^{-9}$, respectively. 
The systematic error due to the slicing procedure is taken as
\bq
 \max\left( \left|I_9-I_7\right|, \left|I_9-I_5\right|, \left|I_7-I_5\right| \right)
\eq
and added linearly to the statistical error. This defines the total error, which is quoted in all results.

The results of this paper are the perturbative coefficients
$A_{{\cal O},n}$, $B_{{\cal O},n}$ and $C_{{\cal O},n}$ for the first ten moments
of the event shape variables
thrust, heavy jet mass, wide jet broadening, total jet broadening, $C$-parameter and the three-to-two jet transition variable.
Since the experimentalists use the heavy jet mass as well as the square root of the heavy jet mass, the results for the moments of both are
given for convenience.
Of course, the $2n$-th moment of $\sqrt{\rho}$ is identical to the $n$-th moment of $\rho$.
The perturbative coefficients for the first ten moments
are tabulated in tables~\ref{table_moments_thrust} to \ref{table_moments_sqrtheavyjetmass}.
The total error is indicated in these tables.

The NNLO coefficient $C_{{\cal O},n}$ can be split up into 
the contributions from the individual colour factors
\bq
 \left. C_{\cal O} \right|_{lc},
 \;\;\;
 \left. C_{\cal O} \right|_{sc},
 \;\;\;
 \left. C_{\cal O} \right|_{ssc},
 \;\;\;
 \left. C_{\cal O} \right|_{nf},
 \;\;\;
 \left. C_{\cal O} \right|_{nfsc},
 \;\;\;
 \left. C_{\cal O} \right|_{nfnf}
\eq
defined in eq.~(\ref{def_colour_factors}).
For the moments the contributions from the individual colour factors to $C_{{\cal O},n}$ are given
in tables~\ref{table_moments_thrust_colour} to \ref{table_moments_sqrtheavyjetmass_colour}.
Again, the total error is indicated in these tables.

In eq.~(\ref{split_up}) we have split up the integral defining the moments into two pieces.
The integral over the region $[\kappa,1]$
\bq
\left\langle {\cal O}^n \right\rangle_{incomplete} & = & 
 \frac{1}{\sigma_{tot}} \int\limits_{\kappa}^1 {\cal O}^n \frac{d\sigma}{d{\cal O}} d{\cal O}
\eq 
is like the complete moment a measurable quantity.
It has the advantage that the perturbative calculation of this quantity is much less 
affected by the numerical issues related to the close-to-two-jet region.
For this reason the results for these ``incomplete'' moments are provided as well.
The perturbative coefficients are given for $\kappa=10^{-3}$ for the various event shape variables in 
tables~\ref{table_moments_thrust_cut} to \ref{table_moments_sqrtheavyjetmass_cut}.
The indicated error corresponds to the statistical error.

In a recent calculation the logarithmic terms of the NNLO coefficient of the thrust distribution have been 
calculated based on soft-collinear effective theory \cite{Becher:2008cf}:
\bq
\label{log_terms}
 \frac{dC_{\tau}}{d\tau}
 & = &
 \frac{1}{\tau} \left[ a_5 \ln^5\tau + a_4 \ln^4\tau + a_3 \ln^3\tau + a_2 \ln^2\tau + a_1 \ln\tau + a_0 + {\cal O}(\tau) \right],
 \;\;\;
 \tau = 1-T.
\eq
The values of the $a_j$'s are for $N_f=5$
\bq
& &
a_5 = -18.96,
 \;\;\;
a_4 = -207.4,
 \;\;\;
a_3 = -122.3,
 \nonumber \\
 & &
a_2 = 1488.3,
 \;\;\;
a_1 = -822.3,
 \;\;\;
a_0 = -683.4.
\eq
The logarithmic terms give a good description of the thrust distribution in the close-to-two jet region.
We may therefore use eq.~(\ref{log_terms}) to estimate the NNLO contribution to the integral 
\bq
 \frac{1}{\sigma_{tot}} \int\limits_0^{\kappa} (1-T)^n \frac{d\sigma}{d(1-T)} d(1-T)
\eq
These results are reported for $\kappa=10^{-3}$, $10^{-5}$, $10^{-7}$ and $10^{-9}$ in table~\ref{table_log_terms}.

The first five moments of the six event shape variables have also been
calculated independently in ref.~\cite{GehrmannDeRidder:2009dp}.
The results of tables~\ref{table_moments_thrust} to \ref{table_moments_y23} for the perturbative coefficients 
$A_{{\cal O},n}$, $B_{{\cal O},n}$ and $C_{{\cal O},n}$ have their correspondence in table 3 of ref.~\cite{GehrmannDeRidder:2009dp}.
The results of tables~\ref{table_moments_thrust_colour} to \ref{table_moments_y23_colour}
for the individual colour factors  to $C_{{\cal O},n}$
have their correspondence in tables 1 and 2 of ref.~\cite{GehrmannDeRidder:2009dp}.
The compatibility of the two results is not perfect, but acceptable given the complexity of the calculation.
The higher moments ($n\ge2$) agree typically within $5\%$.
The agreement in the individual colour factors is in general better (at the level of $2\%$ for the numerically dominant
colour factors $N_c^2$ and $N_f N_c$ for $n\ge2$).
There is a simple explanation for the fact that one observes
a better agreement for the individual colour factors as compared to the complete 
NNLO coefficient $C_{{\cal O},n}$:
The complete NNLO coefficient involves large cancellations between different colour factors 
(in particular between the numerically dominant colour factors $N_c^2$ and $N_f N_c$).
The numerically not so important colour factors $N_c^0$ and $N_c^{-2}$ show are slightly worse agreement.
As they give numerically small contributions they have been calculated with lower statistics.
The agreement for the complete NNLO coefficients for the first moment ($n=1$) is at the level of $20\%$ for
thrust, $C$-parameter and the three-to-two jet transition variable, $50\%$ for the heavy jet mass and
$70\%$ - $100\%$ for the jet broadenings.
It should be noted that the first moment receives sizeable contributions from the close-to-two-jet region,
in particular the jet broadenings.
In comparing the two results one should take into account that tables 1-3 of ref.~\cite{GehrmannDeRidder:2009dp} 
do not include systematic errors.
For the first moment the systematic error from the close-to-two-jet region dominates over the statistical error.
As in the case for the higher moments one also observes for the first moment a better agreement for the individual colour factors
as compared to the complete NNLO results, with the same explanation of large cancellations between different colour factors as above.

\section{Conclusions}
\label{sec:conclusions}

In this article I reported on the NNLO calculation of the moments of the 
event shape observables associated to three-jet events
in electron-positron annihilation.
I provided NNLO results for the moments of the event shape variables
thrust, heavy jet mass, wide jet broadening,
total jet broadening, C parameter and the Durham three-to-two jet transition variable.
The results of this paper will be useful for an extraction of $\alpha_s$ from three-jet 
quantities \cite{Dissertori:2007xa,Gehrmann:2008kh,Dissertori:2009ik,Bethke:2008hf,Pahl:2008uc,Pahl:2009aa}.

\subsection*{Acknowledgements}

I would like to thank Th.~Gehrmann, G.~Heinrich, S. Kluth, Ch.~Pahl and J.~Schieck
for useful discussions.
The computer support from the Max-Planck-Institut for Physics is greatly acknowledged.

\bibliography{/home/stefanw/notes/biblio}
\bibliographystyle{/home/stefanw/latex-style/h-physrev3}

%
%
\clearpage
\begin{table}[p]
\begin{center}
{\scriptsize
\begin{tabular}{|r|c|c|c|}
\hline
 $n$ & $A_{(1-T),n}$ & $B_{(1-T),n}$ & $C_{(1-T),n}$ \\
\hline 
$1$ & $2.10344(3) \cdot 10^{0}$ & $4.499(5) \cdot 10^{1}$ & $1.10(3) \cdot 10^{3}$ \\
$2$ & $1.90190(5) \cdot 10^{-1}$ & $6.2568(4) \cdot 10^{0}$ & $1.829(2) \cdot 10^{2}$ \\
$3$ & $2.9874(1) \cdot 10^{-2}$ & $1.1278(1) \cdot 10^{0}$ & $3.617(3) \cdot 10^{1}$ \\
$4$ & $5.8576(3) \cdot 10^{-3}$ & $2.4619(5) \cdot 10^{-1}$ & $8.447(8) \cdot 10^{0}$ \\
$5$ & $1.29460(9) \cdot 10^{-3}$ & $6.003(2) \cdot 10^{-2}$ & $2.184(2) \cdot 10^{0}$ \\
$6$ & $3.0839(3) \cdot 10^{-4}$ & $1.5709(6) \cdot 10^{-2}$ & $6.026(7) \cdot 10^{-1}$ \\
$7$ & $7.7332(8) \cdot 10^{-5}$ & $4.318(2) \cdot 10^{-3}$ & $1.739(2) \cdot 10^{-1}$ \\
$8$ & $2.0132(2) \cdot 10^{-5}$ & $1.2307(8) \cdot 10^{-3}$ & $5.184(8) \cdot 10^{-2}$ \\
$9$ & $5.3929(7) \cdot 10^{-6}$ & $3.608(3) \cdot 10^{-4}$ & $1.584(3) \cdot 10^{-2}$ \\
$10$ & $1.4775(2) \cdot 10^{-6}$ & $1.0821(9) \cdot 10^{-4}$ & $4.939(9) \cdot 10^{-3}$ \\
\hline
\end{tabular}
}
\caption{\label{table_moments_thrust}
Coefficients of the leading-order ($A_{(1-T),n}$), 
next-to-leading-order ($B_{(1-T),n}$)
and next-to-next-to-leading order ($C_{(1-T),n}$)
contributions to the $n$-th moment of the thrust distribution.
}
\end{center}
\end{table}
\begin{table}[p]
\begin{center}
{\scriptsize
\begin{tabular}{|r|c|c|c|}
\hline
 $n$ & $A_{\rho,n}$ & $B_{\rho,n}$ & $C_{\rho,n}$ \\
\hline 
$1$ & $2.10344(3) \cdot 10^{0}$ & $2.331(4) \cdot 10^{1}$ & $4.2(3) \cdot 10^{2}$ \\
$2$ & $1.90190(5) \cdot 10^{-1}$ & $3.087(1) \cdot 10^{0}$ & $4.15(2) \cdot 10^{1}$ \\
$3$ & $2.9874(1) \cdot 10^{-2}$ & $4.572(3) \cdot 10^{-1}$ & $4.53(4) \cdot 10^{0}$ \\
$4$ & $5.8576(3) \cdot 10^{-3}$ & $8.350(7) \cdot 10^{-2}$ & $5.01(9) \cdot 10^{-1}$ \\
$5$ & $1.29460(9) \cdot 10^{-3}$ & $1.755(2) \cdot 10^{-2}$ & $4.3(3) \cdot 10^{-2}$ \\
$6$ & $3.0839(3) \cdot 10^{-4}$ & $4.082(5) \cdot 10^{-3}$ & $-1.8(8) \cdot 10^{-3}$ \\
$7$ & $7.7332(8) \cdot 10^{-5}$ & $1.026(2) \cdot 10^{-3}$ & $-2.6(2) \cdot 10^{-3}$ \\
$8$ & $2.0132(2) \cdot 10^{-5}$ & $2.744(5) \cdot 10^{-4}$ & $-1.06(8) \cdot 10^{-3}$ \\
$9$ & $5.3929(7) \cdot 10^{-6}$ & $7.71(1) \cdot 10^{-5}$ & $-3.4(3) \cdot 10^{-4}$ \\
$10$ & $1.4775(2) \cdot 10^{-6}$ & $2.257(5) \cdot 10^{-5}$ & $-1.00(8) \cdot 10^{-4}$ \\
\hline
\end{tabular}
}
\caption{\label{table_moments_heavyjetmass}
Coefficients of the leading-order ($A_{\rho,n}$), 
next-to-leading-order ($B_{\rho,n}$)
and next-to-next-to-leading order ($C_{\rho,n}$)
contributions to the $n$-th moment of the heavy jet mass distribution.
}
\end{center}
\end{table}
\begin{table}[p]
\begin{center}
{\scriptsize
\begin{tabular}{|r|c|c|c|}
\hline
 $n$ & $A_{B_W,n}$ & $B_{B_W,n}$ & $C_{B_W,n}$ \\
\hline 
$1$ & $4.067(4) \cdot 10^{0}$ & $-1.01(9) \cdot 10^{1}$ & $2(1) \cdot 10^{3}$ \\
$2$ & $3.36874(5) \cdot 10^{-1}$ & $4.533(1) \cdot 10^{0}$ & $5.09(4) \cdot 10^{1}$ \\
$3$ & $4.7552(1) \cdot 10^{-2}$ & $6.667(2) \cdot 10^{-1}$ & $6.24(3) \cdot 10^{0}$ \\
$4$ & $8.3108(3) \cdot 10^{-3}$ & $1.0681(4) \cdot 10^{-1}$ & $5.99(7) \cdot 10^{-1}$ \\
$5$ & $1.62954(7) \cdot 10^{-3}$ & $1.863(1) \cdot 10^{-2}$ & $1.2(2) \cdot 10^{-2}$ \\
$6$ & $3.4328(2) \cdot 10^{-4}$ & $3.458(3) \cdot 10^{-3}$ & $-1.82(4) \cdot 10^{-2}$ \\
$7$ & $7.5946(4) \cdot 10^{-5}$ & $6.711(7) \cdot 10^{-4}$ & $-8.2(1) \cdot 10^{-3}$ \\
$8$ & $1.7410(1) \cdot 10^{-5}$ & $1.347(2) \cdot 10^{-4}$ & $-2.74(3) \cdot 10^{-3}$ \\
$9$ & $4.1002(3) \cdot 10^{-6}$ & $2.773(5) \cdot 10^{-5}$ & $-8.28(8) \cdot 10^{-4}$ \\
$10$ & $9.8630(8) \cdot 10^{-7}$ & $5.83(1) \cdot 10^{-6}$ & $-2.39(2) \cdot 10^{-4}$ \\
\hline
\end{tabular}
}
\caption{\label{table_moments_widejetbroadening}
Coefficients of the leading-order ($A_{B_W,n}$), 
next-to-leading-order ($B_{B_W,n}$)
and next-to-next-to-leading order ($C_{B_W,n}$)
contributions to the $n$-th moment of the wide jet broadening distribution.
}
\end{center}
\end{table}
\begin{table}[p]
\begin{center}
{\scriptsize
\begin{tabular}{|r|c|c|c|}
\hline
 $n$ & $A_{B_T,n}$ & $B_{B_T,n}$ & $C_{B_T,n}$ \\
\hline 
$1$ & $4.067(4) \cdot 10^{0}$ & $6.41(3) \cdot 10^{1}$ & $2.3(9) \cdot 10^{3}$ \\
$2$ & $3.36874(5) \cdot 10^{-1}$ & $1.473(2) \cdot 10^{1}$ & $3.325(4) \cdot 10^{2}$ \\
$3$ & $4.7552(1) \cdot 10^{-2}$ & $2.766(2) \cdot 10^{0}$ & $7.202(4) \cdot 10^{1}$ \\
$4$ & $8.3108(3) \cdot 10^{-3}$ & $6.070(5) \cdot 10^{-1}$ & $1.676(1) \cdot 10^{1}$ \\
$5$ & $1.62954(7) \cdot 10^{-3}$ & $1.472(1) \cdot 10^{-1}$ & $4.232(3) \cdot 10^{0}$ \\
$6$ & $3.4328(2) \cdot 10^{-4}$ & $3.810(3) \cdot 10^{-2}$ & $1.135(1) \cdot 10^{0}$ \\
$7$ & $7.5946(4) \cdot 10^{-5}$ & $1.0326(7) \cdot 10^{-2}$ & $3.181(3) \cdot 10^{-1}$ \\
$8$ & $1.7410(1) \cdot 10^{-5}$ & $2.897(2) \cdot 10^{-3}$ & $9.21(1) \cdot 10^{-2}$ \\
$9$ & $4.1002(3) \cdot 10^{-6}$ & $8.349(5) \cdot 10^{-4}$ & $2.737(3) \cdot 10^{-2}$ \\
$10$ & $9.8630(8) \cdot 10^{-7}$ & $2.461(1) \cdot 10^{-4}$ & $8.30(1) \cdot 10^{-3}$ \\
\hline
\end{tabular}
}
\caption{\label{table_moments_totaljetbroadening}
Coefficients of the leading-order ($A_{B_T,n}$), 
next-to-leading-order ($B_{B_T,n}$)
and next-to-next-to-leading order ($C_{B_T,n}$)
contributions to the $n$-th moment of the total jet broadening distribution.
}
\end{center}
\end{table}
\begin{table}[p]
\begin{center}
{\scriptsize
\begin{tabular}{|r|c|c|c|}
\hline
 $n$ & $A_{C,n}$ & $B_{C,n}$ & $C_{C,n}$ \\
\hline 
$1$ & $8.6378(1) \cdot 10^{0}$ & $1.728(3) \cdot 10^{2}$ & $4.2(1) \cdot 10^{3}$ \\
$2$ & $2.43160(4) \cdot 10^{0}$ & $8.1160(5) \cdot 10^{1}$ & $2.332(2) \cdot 10^{3}$ \\
$3$ & $1.07919(3) \cdot 10^{0}$ & $4.2752(3) \cdot 10^{1}$ & $1.3608(8) \cdot 10^{3}$ \\
$4$ & $5.6848(2) \cdot 10^{-1}$ & $2.5804(3) \cdot 10^{1}$ & $8.791(5) \cdot 10^{2}$ \\
$5$ & $3.2720(1) \cdot 10^{-1}$ & $1.6865(3) \cdot 10^{1}$ & $6.074(4) \cdot 10^{2}$ \\
$6$ & $1.98719(9) \cdot 10^{-1}$ & $1.1595(2) \cdot 10^{1}$ & $4.386(3) \cdot 10^{2}$ \\
$7$ & $1.25095(6) \cdot 10^{-1}$ & $8.256(2) \cdot 10^{0}$ & $3.265(2) \cdot 10^{2}$ \\
$8$ & $8.0789(4) \cdot 10^{-2}$ & $6.033(2) \cdot 10^{0}$ & $2.485(2) \cdot 10^{2}$ \\
$9$ & $5.3184(3) \cdot 10^{-2}$ & $4.500(1) \cdot 10^{0}$ & $1.925(2) \cdot 10^{2}$ \\
$10$ & $3.5534(2) \cdot 10^{-2}$ & $3.413(1) \cdot 10^{0}$ & $1.511(1) \cdot 10^{2}$ \\
\hline
\end{tabular}
}
\caption{\label{table_moments_Cparameter}
Coefficients of the leading-order ($A_{C,n}$), 
next-to-leading-order ($B_{C,n}$)
and next-to-next-to-leading order ($C_{C,n}$)
contributions to the $n$-th moment of the C-parameter distribution.
}
\end{center}
\end{table}
\begin{table}[p]
\begin{center}
{\scriptsize
\begin{tabular}{|r|c|c|c|}
\hline
 $n$ & $A_{y_{23},n}$ & $B_{y_{23},n}$ & $C_{y_{23},n}$ \\
\hline 
$1$ & $8.9419(2) \cdot 10^{-1}$ & $1.2648(2) \cdot 10^{1}$ & $1.32(2) \cdot 10^{2}$ \\
$2$ & $8.1408(3) \cdot 10^{-2}$ & $1.2912(3) \cdot 10^{0}$ & $1.353(9) \cdot 10^{1}$ \\
$3$ & $1.28529(9) \cdot 10^{-2}$ & $1.9873(9) \cdot 10^{-1}$ & $1.86(2) \cdot 10^{0}$ \\
$4$ & $2.5226(2) \cdot 10^{-3}$ & $3.770(3) \cdot 10^{-2}$ & $3.08(6) \cdot 10^{-1}$ \\
$5$ & $5.5680(6) \cdot 10^{-4}$ & $8.028(8) \cdot 10^{-3}$ & $5.6(2) \cdot 10^{-2}$ \\
$6$ & $1.3232(2) \cdot 10^{-4}$ & $1.837(3) \cdot 10^{-3}$ & $1.03(5) \cdot 10^{-2}$ \\
$7$ & $3.3091(5) \cdot 10^{-5}$ & $4.410(9) \cdot 10^{-4}$ & $1.8(1) \cdot 10^{-3}$ \\
$8$ & $8.590(2) \cdot 10^{-6}$ & $1.096(3) \cdot 10^{-4}$ & $2.8(4) \cdot 10^{-4}$ \\
$9$ & $2.2945(5) \cdot 10^{-6}$ & $2.79(1) \cdot 10^{-5}$ & $2(1) \cdot 10^{-5}$ \\
$10$ & $6.269(1) \cdot 10^{-7}$ & $7.26(3) \cdot 10^{-6}$ & $-9(4) \cdot 10^{-6}$ \\
\hline
\end{tabular}
}
\caption{\label{table_moments_y23}
Coefficients of the leading-order ($A_{y_{23},n}$), 
next-to-leading-order ($B_{y_{23},n}$)
and next-to-next-to-leading order ($C_{y_{23},n}$)
contributions to the $n$-th moment of the three-to-two jet transition distribution.
}
\end{center}
\end{table}
\begin{table}[p]
\begin{center}
{\scriptsize
\begin{tabular}{|r|c|c|c|}
\hline
 $n$ & $A_{\sqrt{\rho},n}$ & $B_{\sqrt{\rho},n}$ & $C_{\sqrt{\rho},n}$ \\
\hline 
$1$ & $1.372(5) \cdot 10^{1}$ & $-3.4(1) \cdot 10^{2}$ & $2(2) \cdot 10^{4}$ \\
$2$ & $2.10344(3) \cdot 10^{0}$ & $2.336(3) \cdot 10^{1}$ & $4.18(6) \cdot 10^{2}$ \\
$3$ & $5.6445(1) \cdot 10^{-1}$ & $8.87(1) \cdot 10^{0}$ & $1.289(4) \cdot 10^{2}$ \\
$4$ & $1.90190(5) \cdot 10^{-1}$ & $3.090(5) \cdot 10^{0}$ & $4.16(2) \cdot 10^{1}$ \\
$5$ & $7.2514(2) \cdot 10^{-2}$ & $1.152(2) \cdot 10^{0}$ & $1.359(7) \cdot 10^{1}$ \\
$6$ & $2.9874(1) \cdot 10^{-2}$ & $4.58(1) \cdot 10^{-1}$ & $4.53(4) \cdot 10^{0}$ \\
$7$ & $1.29764(6) \cdot 10^{-2}$ & $1.915(5) \cdot 10^{-1}$ & $1.52(2) \cdot 10^{0}$ \\
$8$ & $5.8576(3) \cdot 10^{-3}$ & $8.36(3) \cdot 10^{-2}$ & $5.01(9) \cdot 10^{-1}$ \\
$9$ & $2.7224(2) \cdot 10^{-3}$ & $3.78(1) \cdot 10^{-2}$ & $1.57(5) \cdot 10^{-1}$ \\
$10$ & $1.29460(9) \cdot 10^{-3}$ & $1.758(8) \cdot 10^{-2}$ & $4.3(3) \cdot 10^{-2}$ \\
\hline
\end{tabular}
}
\caption{\label{table_moments_sqrtheavyjetmass}
Coefficients of the leading-order ($A_{\sqrt{\rho},n}$), 
next-to-leading-order ($B_{\sqrt{\rho},n}$)
and next-to-next-to-leading order ($C_{\sqrt{\rho},n}$)
contributions to the $n$-th moment of the square root of the heavy jet mass distribution.
}
\end{center}
\end{table}
%

%
%
\clearpage
\begin{table}[p]
\begin{center}
{\scriptsize
\begin{tabular}{|r|c|c|c|}
\hline
 $n$ & $A_{(1-T),n}$ & $B_{(1-T),n}$ & $C_{(1-T),n}$ \\
\hline 
$1$ & $2.06527(3) \cdot 10^{0}$ & $4.8055(3) \cdot 10^{1}$ & $1.076(1) \cdot 10^{3}$ \\
$2$ & $1.90172(5) \cdot 10^{-1}$ & $6.2578(4) \cdot 10^{0}$ & $1.829(1) \cdot 10^{2}$ \\
$3$ & $2.9874(1) \cdot 10^{-2}$ & $1.1278(1) \cdot 10^{0}$ & $3.617(3) \cdot 10^{1}$ \\
$4$ & $5.8576(3) \cdot 10^{-3}$ & $2.4619(3) \cdot 10^{-1}$ & $8.447(8) \cdot 10^{0}$ \\
$5$ & $1.29460(8) \cdot 10^{-3}$ & $6.0029(9) \cdot 10^{-2}$ & $2.184(2) \cdot 10^{0}$ \\
$6$ & $3.0839(2) \cdot 10^{-4}$ & $1.5710(3) \cdot 10^{-2}$ & $6.026(7) \cdot 10^{-1}$ \\
$7$ & $7.7332(7) \cdot 10^{-5}$ & $4.318(1) \cdot 10^{-3}$ & $1.739(2) \cdot 10^{-1}$ \\
$8$ & $2.0132(2) \cdot 10^{-5}$ & $1.2307(3) \cdot 10^{-3}$ & $5.184(8) \cdot 10^{-2}$ \\
$9$ & $5.3929(6) \cdot 10^{-6}$ & $3.608(1) \cdot 10^{-4}$ & $1.584(3) \cdot 10^{-2}$ \\
$10$ & $1.4775(2) \cdot 10^{-6}$ & $1.0821(4) \cdot 10^{-4}$ & $4.939(9) \cdot 10^{-3}$ \\
\hline
\end{tabular}
}
\caption{\label{table_moments_thrust_cut}
Coefficients of the leading-order ($A_{(1-T),n}$), 
next-to-leading-order ($B_{(1-T),n}$)
and next-to-next-to-leading order ($C_{(1-T),n}$)
contributions to the $n$-th moment of the thrust distribution with a cut $(1-T)>\kappa$ for $\kappa=10^{-3}$.
}
\end{center}
\end{table}
\begin{table}[p]
\begin{center}
{\scriptsize
\begin{tabular}{|r|c|c|c|}
\hline
 $n$ & $A_{\rho,n}$ & $B_{\rho,n}$ & $C_{\rho,n}$ \\
\hline 
$1$ & $2.06527(3) \cdot 10^{0}$ & $2.6555(3) \cdot 10^{1}$ & $3.55(1) \cdot 10^{2}$ \\
$2$ & $1.90172(5) \cdot 10^{-1}$ & $3.0879(5) \cdot 10^{0}$ & $4.16(2) \cdot 10^{1}$ \\
$3$ & $2.9874(1) \cdot 10^{-2}$ & $4.572(1) \cdot 10^{-1}$ & $4.53(4) \cdot 10^{0}$ \\
$4$ & $5.8576(3) \cdot 10^{-3}$ & $8.350(3) \cdot 10^{-2}$ & $5.01(9) \cdot 10^{-1}$ \\
$5$ & $1.29460(8) \cdot 10^{-3}$ & $1.7550(9) \cdot 10^{-2}$ & $4.3(3) \cdot 10^{-2}$ \\
$6$ & $3.0839(2) \cdot 10^{-4}$ & $4.082(3) \cdot 10^{-3}$ & $-1.8(8) \cdot 10^{-3}$ \\
$7$ & $7.7332(7) \cdot 10^{-5}$ & $1.0262(9) \cdot 10^{-3}$ & $-2.6(2) \cdot 10^{-3}$ \\
$8$ & $2.0132(2) \cdot 10^{-5}$ & $2.744(3) \cdot 10^{-4}$ & $-1.06(8) \cdot 10^{-3}$ \\
$9$ & $5.3929(6) \cdot 10^{-6}$ & $7.711(9) \cdot 10^{-5}$ & $-3.4(3) \cdot 10^{-4}$ \\
$10$ & $1.4775(2) \cdot 10^{-6}$ & $2.257(3) \cdot 10^{-5}$ & $-1.00(8) \cdot 10^{-4}$ \\
\hline
\end{tabular}
}
\caption{\label{table_moments_heavyjetmass_cut}
Coefficients of the leading-order ($A_{\rho,n}$), 
next-to-leading-order ($B_{\rho,n}$)
and next-to-next-to-leading order ($C_{\rho,n}$)
contributions to the $n$-th moment of the heavy jet mass distribution with a cut $\rho>\kappa$ for $\kappa=10^{-3}$.
}
\end{center}
\end{table}
\begin{table}[p]
\begin{center}
{\scriptsize
\begin{tabular}{|r|c|c|c|}
\hline
 $n$ & $A_{B_W,n}$ & $B_{B_W,n}$ & $C_{B_W,n}$ \\
\hline 
$1$ & $3.99153(5) \cdot 10^{0}$ & $5.929(8) \cdot 10^{0}$ & $3.56(4) \cdot 10^{2}$ \\
$2$ & $3.36839(5) \cdot 10^{-1}$ & $4.5391(5) \cdot 10^{0}$ & $5.06(2) \cdot 10^{1}$ \\
$3$ & $4.7552(1) \cdot 10^{-2}$ & $6.667(1) \cdot 10^{-1}$ & $6.24(3) \cdot 10^{0}$ \\
$4$ & $8.3108(3) \cdot 10^{-3}$ & $1.0680(2) \cdot 10^{-1}$ & $5.99(7) \cdot 10^{-1}$ \\
$5$ & $1.62954(7) \cdot 10^{-3}$ & $1.8634(6) \cdot 10^{-2}$ & $1.2(2) \cdot 10^{-2}$ \\
$6$ & $3.4328(2) \cdot 10^{-4}$ & $3.458(2) \cdot 10^{-3}$ & $-1.82(4) \cdot 10^{-2}$ \\
$7$ & $7.5946(4) \cdot 10^{-5}$ & $6.711(4) \cdot 10^{-4}$ & $-8.2(1) \cdot 10^{-3}$ \\
$8$ & $1.7410(1) \cdot 10^{-5}$ & $1.347(1) \cdot 10^{-4}$ & $-2.74(3) \cdot 10^{-3}$ \\
$9$ & $4.1002(3) \cdot 10^{-6}$ & $2.773(3) \cdot 10^{-5}$ & $-8.28(8) \cdot 10^{-4}$ \\
$10$ & $9.8630(7) \cdot 10^{-7}$ & $5.831(8) \cdot 10^{-6}$ & $-2.39(2) \cdot 10^{-4}$ \\
\hline
\end{tabular}
}
\caption{\label{table_moments_widejetbroadening_cut}
Coefficients of the leading-order ($A_{B_W,n}$), 
next-to-leading-order ($B_{B_W,n}$)
and next-to-next-to-leading order ($C_{B_W,n}$)
contributions to the $n$-th moment of the wide jet broadening distribution with a cut $B_W>\kappa$ for $\kappa=10^{-3}$.
}
\end{center}
\end{table}
\begin{table}[p]
\begin{center}
{\scriptsize
\begin{tabular}{|r|c|c|c|}
\hline
 $n$ & $A_{B_T,n}$ & $B_{B_T,n}$ & $C_{B_T,n}$ \\
\hline 
$1$ & $3.99153(5) \cdot 10^{0}$ & $7.9498(8) \cdot 10^{1}$ & $9.72(4) \cdot 10^{2}$ \\
$2$ & $3.36839(5) \cdot 10^{-1}$ & $1.47365(7) \cdot 10^{1}$ & $3.324(2) \cdot 10^{2}$ \\
$3$ & $4.7552(1) \cdot 10^{-2}$ & $2.7659(2) \cdot 10^{0}$ & $7.202(4) \cdot 10^{1}$ \\
$4$ & $8.3108(3) \cdot 10^{-3}$ & $6.0697(5) \cdot 10^{-1}$ & $1.676(1) \cdot 10^{1}$ \\
$5$ & $1.62954(7) \cdot 10^{-3}$ & $1.4718(2) \cdot 10^{-1}$ & $4.232(3) \cdot 10^{0}$ \\
$6$ & $3.4328(2) \cdot 10^{-4}$ & $3.8099(5) \cdot 10^{-2}$ & $1.135(1) \cdot 10^{0}$ \\
$7$ & $7.5946(4) \cdot 10^{-5}$ & $1.0326(2) \cdot 10^{-2}$ & $3.181(3) \cdot 10^{-1}$ \\
$8$ & $1.7410(1) \cdot 10^{-5}$ & $2.8966(6) \cdot 10^{-3}$ & $9.21(1) \cdot 10^{-2}$ \\
$9$ & $4.1002(3) \cdot 10^{-6}$ & $8.349(2) \cdot 10^{-4}$ & $2.737(3) \cdot 10^{-2}$ \\
$10$ & $9.8630(7) \cdot 10^{-7}$ & $2.4608(6) \cdot 10^{-4}$ & $8.30(1) \cdot 10^{-3}$ \\
\hline
\end{tabular}
}
\caption{\label{table_moments_totaljetbroadening_cut}
Coefficients of the leading-order ($A_{B_T,n}$), 
next-to-leading-order ($B_{B_T,n}$)
and next-to-next-to-leading order ($C_{B_T,n}$)
contributions to the $n$-th moment of the total jet broadening distribution with a cut $B_T>\kappa$ for $\kappa=10^{-3}$.
}
\end{center}
\end{table}
\begin{table}[p]
\begin{center}
{\scriptsize
\begin{tabular}{|r|c|c|c|}
\hline
 $n$ & $A_{C,n}$ & $B_{C,n}$ & $C_{C,n}$ \\
\hline 
$1$ & $8.59006(9) \cdot 10^{0}$ & $1.7964(1) \cdot 10^{2}$ & $3.930(6) \cdot 10^{3}$ \\
$2$ & $2.43158(4) \cdot 10^{0}$ & $8.1163(4) \cdot 10^{1}$ & $2.332(1) \cdot 10^{3}$ \\
$3$ & $1.07919(3) \cdot 10^{0}$ & $4.2752(3) \cdot 10^{1}$ & $1.3608(8) \cdot 10^{3}$ \\
$4$ & $5.6848(2) \cdot 10^{-1}$ & $2.5804(2) \cdot 10^{1}$ & $8.791(5) \cdot 10^{2}$ \\
$5$ & $3.2720(1) \cdot 10^{-1}$ & $1.6865(2) \cdot 10^{1}$ & $6.074(4) \cdot 10^{2}$ \\
$6$ & $1.98719(9) \cdot 10^{-1}$ & $1.1595(1) \cdot 10^{1}$ & $4.386(3) \cdot 10^{2}$ \\
$7$ & $1.25095(6) \cdot 10^{-1}$ & $8.256(1) \cdot 10^{0}$ & $3.265(2) \cdot 10^{2}$ \\
$8$ & $8.0789(4) \cdot 10^{-2}$ & $6.033(1) \cdot 10^{0}$ & $2.485(2) \cdot 10^{2}$ \\
$9$ & $5.3184(3) \cdot 10^{-2}$ & $4.4999(8) \cdot 10^{0}$ & $1.925(2) \cdot 10^{2}$ \\
$10$ & $3.5534(2) \cdot 10^{-2}$ & $3.4129(7) \cdot 10^{0}$ & $1.511(1) \cdot 10^{2}$ \\
\hline
\end{tabular}
}
\caption{\label{table_moments_Cparameter_cut}
Coefficients of the leading-order ($A_{C,n}$), 
next-to-leading-order ($B_{C,n}$)
and next-to-next-to-leading order ($C_{C,n}$)
contributions to the $n$-th moment of the C-parameter distribution with a cut $C>\kappa$ for $\kappa=10^{-3}$.
}
\end{center}
\end{table}
\begin{table}[p]
\begin{center}
{\scriptsize
\begin{tabular}{|r|c|c|c|}
\hline
 $n$ & $A_{y_{23},n}$ & $B_{y_{23},n}$ & $C_{y_{23},n}$ \\
\hline 
$1$ & $8.7701(2) \cdot 10^{-1}$ & $1.3044(2) \cdot 10^{1}$ & $1.367(6) \cdot 10^{2}$ \\
$2$ & $8.1400(3) \cdot 10^{-2}$ & $1.2913(3) \cdot 10^{0}$ & $1.353(9) \cdot 10^{1}$ \\
$3$ & $1.28529(8) \cdot 10^{-2}$ & $1.9873(7) \cdot 10^{-1}$ & $1.86(2) \cdot 10^{0}$ \\
$4$ & $2.5226(2) \cdot 10^{-3}$ & $3.770(2) \cdot 10^{-2}$ & $3.08(6) \cdot 10^{-1}$ \\
$5$ & $5.5680(6) \cdot 10^{-4}$ & $8.028(6) \cdot 10^{-3}$ & $5.6(2) \cdot 10^{-2}$ \\
$6$ & $1.3232(2) \cdot 10^{-4}$ & $1.837(2) \cdot 10^{-3}$ & $1.03(5) \cdot 10^{-2}$ \\
$7$ & $3.3091(4) \cdot 10^{-5}$ & $4.410(6) \cdot 10^{-4}$ & $1.8(1) \cdot 10^{-3}$ \\
$8$ & $8.590(1) \cdot 10^{-6}$ & $1.096(2) \cdot 10^{-4}$ & $2.8(4) \cdot 10^{-4}$ \\
$9$ & $2.2945(4) \cdot 10^{-6}$ & $2.794(5) \cdot 10^{-5}$ & $2(1) \cdot 10^{-5}$ \\
$10$ & $6.269(1) \cdot 10^{-7}$ & $7.26(2) \cdot 10^{-6}$ & $-9(4) \cdot 10^{-6}$ \\
\hline
\end{tabular}
}
\caption{\label{table_moments_y23_cut}
Coefficients of the leading-order ($A_{y_{23},n}$), 
next-to-leading-order ($B_{y_{23},n}$)
and next-to-next-to-leading order ($C_{y_{23},n}$)
contributions to the $n$-th moment of the three-to-two jet transition distribution with a cut $y_{23}>\kappa$ for $\kappa=10^{-3}$.
}
\end{center}
\end{table}
\begin{table}[p]
\begin{center}
{\scriptsize
\begin{tabular}{|r|c|c|c|}
\hline
 $n$ & $A_{\sqrt{\rho},n}$ & $B_{\sqrt{\rho},n}$ & $C_{\sqrt{\rho},n}$ \\
\hline 
$1$ & $1.35675(6) \cdot 10^{1}$ & $-2.714(2) \cdot 10^{2}$ & $8.86(4) \cdot 10^{3}$ \\
$2$ & $2.10336(3) \cdot 10^{0}$ & $2.3347(3) \cdot 10^{1}$ & $4.12(1) \cdot 10^{2}$ \\
$3$ & $5.6445(1) \cdot 10^{-1}$ & $8.857(1) \cdot 10^{0}$ & $1.289(4) \cdot 10^{2}$ \\
$4$ & $1.90190(5) \cdot 10^{-1}$ & $3.0867(5) \cdot 10^{0}$ & $4.16(2) \cdot 10^{1}$ \\
$5$ & $7.2514(2) \cdot 10^{-2}$ & $1.1506(2) \cdot 10^{0}$ & $1.359(7) \cdot 10^{1}$ \\
$6$ & $2.9874(1) \cdot 10^{-2}$ & $4.572(1) \cdot 10^{-1}$ & $4.53(4) \cdot 10^{0}$ \\
$7$ & $1.29764(6) \cdot 10^{-2}$ & $1.9132(6) \cdot 10^{-1}$ & $1.52(2) \cdot 10^{0}$ \\
$8$ & $5.8576(3) \cdot 10^{-3}$ & $8.350(3) \cdot 10^{-2}$ & $5.01(9) \cdot 10^{-1}$ \\
$9$ & $2.7224(2) \cdot 10^{-3}$ & $3.773(2) \cdot 10^{-2}$ & $1.57(5) \cdot 10^{-1}$ \\
$10$ & $1.29460(8) \cdot 10^{-3}$ & $1.7550(9) \cdot 10^{-2}$ & $4.3(3) \cdot 10^{-2}$ \\
\hline
\end{tabular}
}
\caption{\label{table_moments_sqrtheavyjetmass_cut}
Coefficients of the leading-order ($A_{\sqrt{\rho},n}$), 
next-to-leading-order ($B_{\sqrt{\rho},n}$)
and next-to-next-to-leading order ($C_{\sqrt{\rho},n}$)
contributions to the $n$-th moment of the square root of the heavy jet mass distribution with a cut $\sqrt{\rho}>\kappa$ for $\kappa=10^{-3}$.
}
\end{center}
\end{table}
%

\clearpage
\begin{table}[p]
\begin{center}
{\scriptsize
\begin{tabular}{|r|c|c|c|c|c|c|}
\hline
 $n$ & $N_c^2$ & $N_c^0$ & $N_c^{-2}$ & $N_f N_c$ & $N_f/N_c$ & $N_f^2$ \\
\hline 
$1$ & $2.51(1) \cdot 10^{3}$ & $3.7(5) \cdot 10^{1}$ & $6(5) \cdot 10^{-1}$ & $-1.69(3) \cdot 10^{3}$ & $-1.2(5) \cdot 10^{1}$ & $2.55(5) \cdot 10^{2}$ \\
$2$ & $3.460(1) \cdot 10^{2}$ & $-2.003(2) \cdot 10^{1}$ & $-2.293(5) \cdot 10^{-1}$ & $-1.6275(8) \cdot 10^{2}$ & $7.776(4) \cdot 10^{0}$ & $1.2152(3) \cdot 10^{1}$ \\
$3$ & $6.368(3) \cdot 10^{1}$ & $-4.697(4) \cdot 10^{0}$ & $2(7) \cdot 10^{-5}$ & $-2.568(1) \cdot 10^{1}$ & $1.482(1) \cdot 10^{0}$ & $1.3903(4) \cdot 10^{0}$ \\
$4$ & $1.4240(7) \cdot 10^{1}$ & $-1.137(1) \cdot 10^{0}$ & $5.20(2) \cdot 10^{-3}$ & $-5.196(4) \cdot 10^{0}$ & $3.221(3) \cdot 10^{-1}$ & $2.1322(9) \cdot 10^{-1}$ \\
$5$ & $3.563(2) \cdot 10^{0}$ & $-2.956(3) \cdot 10^{-1}$ & $2.137(8) \cdot 10^{-3}$ & $-1.200(1) \cdot 10^{0}$ & $7.751(8) \cdot 10^{-2}$ & $3.678(3) \cdot 10^{-2}$ \\
$6$ & $9.572(7) \cdot 10^{-1}$ & $-8.13(1) \cdot 10^{-2}$ & $7.39(3) \cdot 10^{-4}$ & $-3.006(3) \cdot 10^{-1}$ & $2.000(3) \cdot 10^{-2}$ & $6.503(8) \cdot 10^{-3}$ \\
$7$ & $2.699(2) \cdot 10^{-1}$ & $-2.330(3) \cdot 10^{-2}$ & $2.461(9) \cdot 10^{-4}$ & $-7.95(1) \cdot 10^{-2}$ & $5.423(9) \cdot 10^{-3}$ & $1.070(2) \cdot 10^{-3}$ \\
$8$ & $7.885(7) \cdot 10^{-2}$ & $-6.90(1) \cdot 10^{-3}$ & $8.12(3) \cdot 10^{-5}$ & $-2.185(3) \cdot 10^{-2}$ & $1.526(3) \cdot 10^{-3}$ & $1.275(7) \cdot 10^{-4}$ \\
$9$ & $2.366(2) \cdot 10^{-2}$ & $-2.093(4) \cdot 10^{-3}$ & $2.68(1) \cdot 10^{-5}$ & $-6.18(1) \cdot 10^{-3}$ & $4.416(9) \cdot 10^{-4}$ & $-8.8(2) \cdot 10^{-6}$ \\
$10$ & $7.253(8) \cdot 10^{-3}$ & $-6.48(1) \cdot 10^{-4}$ & $8.91(4) \cdot 10^{-6}$ & $-1.791(3) \cdot 10^{-3}$ & $1.307(3) \cdot 10^{-4}$ & $-1.511(8) \cdot 10^{-5}$ \\
\hline
\end{tabular}
}
\caption{\label{table_moments_thrust_colour}
Individual contributions from the different colour factors to the next-to-next-to-leading order
contribution to the $n$-th moment of the thrust distribution.
}
\end{center}
\end{table}

\begin{table}[p]
\begin{center}
{\scriptsize
\begin{tabular}{|r|c|c|c|c|c|c|}
\hline
 $n$ & $N_c^2$ & $N_c^0$ & $N_c^{-2}$ & $N_f N_c$ & $N_f/N_c$ & $N_f^2$ \\
\hline 
$1$ & $1.23(3) \cdot 10^{3}$ & $9.0(3) \cdot 10^{1}$ & $8(1) \cdot 10^{0}$ & $-1.10(2) \cdot 10^{3}$ & $-7.1(6) \cdot 10^{1}$ & $2.72(5) \cdot 10^{2}$ \\
$2$ & $1.234(1) \cdot 10^{2}$ & $-2.47(2) \cdot 10^{0}$ & $-8.45(4) \cdot 10^{-2}$ & $-9.575(7) \cdot 10^{1}$ & $1.422(4) \cdot 10^{0}$ & $1.5031(3) \cdot 10^{1}$ \\
$3$ & $1.574(3) \cdot 10^{1}$ & $-6.32(4) \cdot 10^{-1}$ & $-1.20(1) \cdot 10^{-2}$ & $-1.290(1) \cdot 10^{1}$ & $3.167(9) \cdot 10^{-1}$ & $2.0142(4) \cdot 10^{0}$ \\
$4$ & $2.458(9) \cdot 10^{0}$ & $-1.13(1) \cdot 10^{-1}$ & $-2.09(3) \cdot 10^{-3}$ & $-2.269(4) \cdot 10^{0}$ & $6.12(2) \cdot 10^{-2}$ & $3.6598(9) \cdot 10^{-1}$ \\
$5$ & $4.40(2) \cdot 10^{-1}$ & $-2.04(3) \cdot 10^{-2}$ & $-4.67(9) \cdot 10^{-4}$ & $-4.66(1) \cdot 10^{-1}$ & $1.299(7) \cdot 10^{-2}$ & $7.700(3) \cdot 10^{-2}$ \\
$6$ & $8.77(7) \cdot 10^{-2}$ & $-3.8(1) \cdot 10^{-3}$ & $-1.22(3) \cdot 10^{-4}$ & $-1.062(3) \cdot 10^{-1}$ & $3.04(2) \cdot 10^{-3}$ & $1.7616(7) \cdot 10^{-2}$ \\
$7$ & $1.92(2) \cdot 10^{-2}$ & $-7.4(3) \cdot 10^{-4}$ & $-3.50(9) \cdot 10^{-5}$ & $-2.607(9) \cdot 10^{-2}$ & $7.73(7) \cdot 10^{-4}$ & $4.248(2) \cdot 10^{-3}$ \\
$8$ & $4.60(7) \cdot 10^{-3}$ & $-1.5(1) \cdot 10^{-4}$ & $-1.06(3) \cdot 10^{-5}$ & $-6.77(3) \cdot 10^{-3}$ & $2.11(2) \cdot 10^{-4}$ & $1.0610(6) \cdot 10^{-3}$ \\
$9$ & $1.20(2) \cdot 10^{-3}$ & $-3.0(3) \cdot 10^{-5}$ & $-3.3(1) \cdot 10^{-6}$ & $-1.839(9) \cdot 10^{-3}$ & $6.11(7) \cdot 10^{-5}$ & $2.713(2) \cdot 10^{-4}$ \\
$10$ & $3.37(8) \cdot 10^{-4}$ & $-6(1) \cdot 10^{-6}$ & $-1.08(3) \cdot 10^{-6}$ & $-5.18(3) \cdot 10^{-4}$ & $1.84(2) \cdot 10^{-5}$ & $7.044(6) \cdot 10^{-5}$ \\
\hline
\end{tabular}
}
\caption{\label{table_moments_heavyjetmass_colour}
Individual contributions from the different colour factors to the next-to-next-to-leading order
contribution to the $n$-th moment of the heavy jet mass distribution.
}
\end{center}
\end{table}

\begin{table}[p]
\begin{center}
{\scriptsize
\begin{tabular}{|r|c|c|c|c|c|c|}
\hline
 $n$ & $N_c^2$ & $N_c^0$ & $N_c^{-2}$ & $N_f N_c$ & $N_f/N_c$ & $N_f^2$ \\
\hline 
$1$ & $2(1) \cdot 10^{3}$ & $0(2) \cdot 10^{1}$ & $8(3) \cdot 10^{1}$ & $-7(3) \cdot 10^{2}$ & $-4.5(7) \cdot 10^{2}$ & $7.3(4) \cdot 10^{2}$ \\
$2$ & $1.846(3) \cdot 10^{2}$ & $6.94(7) \cdot 10^{0}$ & $2.13(8) \cdot 10^{-1}$ & $-1.712(2) \cdot 10^{2}$ & $-3.19(3) \cdot 10^{0}$ & $3.364(2) \cdot 10^{1}$ \\
$3$ & $2.395(3) \cdot 10^{1}$ & $-1.20(4) \cdot 10^{-1}$ & $-1.054(8) \cdot 10^{-2}$ & $-2.155(1) \cdot 10^{1}$ & $1.330(7) \cdot 10^{-1}$ & $3.8438(4) \cdot 10^{0}$ \\
$4$ & $3.287(7) \cdot 10^{0}$ & $-6.55(9) \cdot 10^{-2}$ & $-1.89(2) \cdot 10^{-3}$ & $-3.271(3) \cdot 10^{0}$ & $3.97(2) \cdot 10^{-2}$ & $6.1019(9) \cdot 10^{-1}$ \\
$5$ & $4.65(2) \cdot 10^{-1}$ & $-1.19(2) \cdot 10^{-2}$ & $-2.92(4) \cdot 10^{-4}$ & $-5.618(7) \cdot 10^{-1}$ & $7.29(4) \cdot 10^{-3}$ & $1.1390(2) \cdot 10^{-1}$ \\
$6$ & $6.35(4) \cdot 10^{-2}$ & $-1.61(5) \cdot 10^{-3}$ & $-4.8(1) \cdot 10^{-5}$ & $-1.046(2) \cdot 10^{-1}$ & $1.21(1) \cdot 10^{-3}$ & $2.3348(5) \cdot 10^{-2}$ \\
$7$ & $7.3(1) \cdot 10^{-3}$ & $-1.1(1) \cdot 10^{-4}$ & $-8.5(3) \cdot 10^{-6}$ & $-2.060(5) \cdot 10^{-2}$ & $1.82(3) \cdot 10^{-4}$ & $5.083(1) \cdot 10^{-3}$ \\
$8$ & $2.8(3) \cdot 10^{-4}$ & $3.2(3) \cdot 10^{-5}$ & $-1.59(9) \cdot 10^{-6}$ & $-4.22(1) \cdot 10^{-3}$ & $2.22(7) \cdot 10^{-5}$ & $1.1533(3) \cdot 10^{-3}$ \\
$9$ & $-2.25(7) \cdot 10^{-4}$ & $2.04(9) \cdot 10^{-5}$ & $-3.1(2) \cdot 10^{-7}$ & $-8.94(3) \cdot 10^{-4}$ & $9(2) \cdot 10^{-7}$ & $2.6970(9) \cdot 10^{-4}$ \\
$10$ & $-1.17(2) \cdot 10^{-4}$ & $7.8(2) \cdot 10^{-6}$ & $-6.5(7) \cdot 10^{-8}$ & $-1.940(9) \cdot 10^{-4}$ & $-7.3(5) \cdot 10^{-7}$ & $6.452(2) \cdot 10^{-5}$ \\
\hline
\end{tabular}
}
\caption{\label{table_moments_widejetbroadening_colour}
Individual contributions from the different colour factors to the next-to-next-to-leading order
contribution to the $n$-th moment of the wide jet broadening distribution.
}
\end{center}
\end{table}

\begin{table}[p]
\begin{center}
{\scriptsize
\begin{tabular}{|r|c|c|c|c|c|c|}
\hline
 $n$ & $N_c^2$ & $N_c^0$ & $N_c^{-2}$ & $N_f N_c$ & $N_f/N_c$ & $N_f^2$ \\
\hline 
$1$ & $4.5(7) \cdot 10^{3}$ & $2(1) \cdot 10^{2}$ & $3(2) \cdot 10^{1}$ & $-3.0(3) \cdot 10^{3}$ & $-2.2(7) \cdot 10^{2}$ & $7.0(4) \cdot 10^{2}$ \\
$2$ & $7.443(3) \cdot 10^{2}$ & $-2.480(6) \cdot 10^{1}$ & $-2.516(7) \cdot 10^{0}$ & $-4.359(2) \cdot 10^{2}$ & $2.356(3) \cdot 10^{1}$ & $2.790(2) \cdot 10^{1}$ \\
$3$ & $1.4437(4) \cdot 10^{2}$ & $-9.048(7) \cdot 10^{0}$ & $-3.009(2) \cdot 10^{-1}$ & $-7.031(2) \cdot 10^{1}$ & $4.884(2) \cdot 10^{0}$ & $2.4263(4) \cdot 10^{0}$ \\
$4$ & $3.198(1) \cdot 10^{1}$ & $-2.354(2) \cdot 10^{0}$ & $-4.384(5) \cdot 10^{-2}$ & $-1.4127(5) \cdot 10^{1}$ & $1.0712(7) \cdot 10^{0}$ & $2.359(1) \cdot 10^{-1}$ \\
$5$ & $7.810(3) \cdot 10^{0}$ & $-6.163(5) \cdot 10^{-1}$ & $-7.22(1) \cdot 10^{-3}$ & $-3.222(2) \cdot 10^{0}$ & $2.560(2) \cdot 10^{-1}$ & $1.100(3) \cdot 10^{-2}$ \\
$6$ & $2.0388(9) \cdot 10^{0}$ & $-1.671(2) \cdot 10^{-1}$ & $-1.225(4) \cdot 10^{-3}$ & $-7.948(5) \cdot 10^{-1}$ & $6.516(6) \cdot 10^{-2}$ & $-5.754(8) \cdot 10^{-3}$ \\
$7$ & $5.579(3) \cdot 10^{-1}$ & $-4.686(5) \cdot 10^{-2}$ & $-1.92(1) \cdot 10^{-4}$ & $-2.068(1) \cdot 10^{-1}$ & $1.737(2) \cdot 10^{-2}$ & $-3.332(3) \cdot 10^{-3}$ \\
$8$ & $1.5813(9) \cdot 10^{-1}$ & $-1.352(2) \cdot 10^{-2}$ & $-2.04(5) \cdot 10^{-5}$ & $-5.592(5) \cdot 10^{-2}$ & $4.797(6) \cdot 10^{-3}$ & $-1.3251(8) \cdot 10^{-3}$ \\
$9$ & $4.605(3) \cdot 10^{-2}$ & $-3.995(5) \cdot 10^{-3}$ & $2.7(2) \cdot 10^{-6}$ & $-1.558(2) \cdot 10^{-2}$ & $1.363(2) \cdot 10^{-3}$ & $-4.721(3) \cdot 10^{-4}$ \\
$10$ & $1.371(1) \cdot 10^{-2}$ & $-1.204(2) \cdot 10^{-3}$ & $3.06(6) \cdot 10^{-6}$ & $-4.446(5) \cdot 10^{-3}$ & $3.961(6) \cdot 10^{-4}$ & $-1.607(1) \cdot 10^{-4}$ \\
\hline
\end{tabular}
}
\caption{\label{table_moments_totaljetbroadening_colour}
Individual contributions from the different colour factors to the next-to-next-to-leading order
contribution to the $n$-th moment of the total jet broadening distribution.
}
\end{center}
\end{table}

\begin{table}[p]
\begin{center}
{\scriptsize
\begin{tabular}{|r|c|c|c|c|c|c|}
\hline
 $n$ & $N_c^2$ & $N_c^0$ & $N_c^{-2}$ & $N_f N_c$ & $N_f/N_c$ & $N_f^2$ \\
\hline 
$1$ & $9.8(1) \cdot 10^{3}$ & $2.9(2) \cdot 10^{2}$ & $7(5) \cdot 10^{0}$ & $-6.87(8) \cdot 10^{3}$ & $-1.1(2) \cdot 10^{2}$ & $1.12(2) \cdot 10^{3}$ \\
$2$ & $4.528(1) \cdot 10^{3}$ & $-2.248(2) \cdot 10^{2}$ & $-5.047(4) \cdot 10^{0}$ & $-2.2420(7) \cdot 10^{3}$ & $1.0303(5) \cdot 10^{2}$ & $1.7314(2) \cdot 10^{2}$ \\
$3$ & $2.4430(7) \cdot 10^{3}$ & $-1.672(1) \cdot 10^{2}$ & $-9.47(2) \cdot 10^{-1}$ & $-1.0292(4) \cdot 10^{3}$ & $5.944(3) \cdot 10^{1}$ & $5.5725(9) \cdot 10^{1}$ \\
$4$ & $1.5070(5) \cdot 10^{3}$ & $-1.1508(8) \cdot 10^{2}$ & $6.1(1) \cdot 10^{-2}$ & $-5.713(2) \cdot 10^{2}$ & $3.628(2) \cdot 10^{1}$ & $2.2186(6) \cdot 10^{1}$ \\
$5$ & $1.0065(3) \cdot 10^{3}$ & $-8.138(6) \cdot 10^{1}$ & $3.29(1) \cdot 10^{-1}$ & $-3.509(2) \cdot 10^{2}$ & $2.362(2) \cdot 10^{1}$ & $9.238(4) \cdot 10^{0}$ \\
$6$ & $7.069(3) \cdot 10^{2}$ & $-5.934(4) \cdot 10^{1}$ & $3.844(9) \cdot 10^{-1}$ & $-2.290(1) \cdot 10^{2}$ & $1.611(1) \cdot 10^{1}$ & $3.512(3) \cdot 10^{0}$ \\
$7$ & $5.140(2) \cdot 10^{2}$ & $-4.437(3) \cdot 10^{1}$ & $3.709(8) \cdot 10^{-1}$ & $-1.557(1) \cdot 10^{2}$ & $1.136(1) \cdot 10^{1}$ & $8.27(2) \cdot 10^{-1}$ \\
$8$ & $3.832(2) \cdot 10^{2}$ & $-3.386(3) \cdot 10^{1}$ & $3.357(7) \cdot 10^{-1}$ & $-1.0894(8) \cdot 10^{2}$ & $8.214(9) \cdot 10^{0}$ & $-4.39(2) \cdot 10^{-1}$ \\
$9$ & $2.913(1) \cdot 10^{2}$ & $-2.627(2) \cdot 10^{1}$ & $2.956(6) \cdot 10^{-1}$ & $-7.793(7) \cdot 10^{1}$ & $6.059(8) \cdot 10^{0}$ & $-1.002(2) \cdot 10^{0}$ \\
$10$ & $2.250(1) \cdot 10^{2}$ & $-2.067(2) \cdot 10^{1}$ & $2.570(5) \cdot 10^{-1}$ & $-5.674(6) \cdot 10^{1}$ & $4.544(6) \cdot 10^{0}$ & $-1.209(1) \cdot 10^{0}$ \\
\hline
\end{tabular}
}
\caption{\label{table_moments_Cparameter_colour}
Individual contributions from the different colour factors to the next-to-next-to-leading order
contribution to the $n$-th moment of the C parameter distribution.
}
\end{center}
\end{table}

\begin{table}[p]
\begin{center}
{\scriptsize
\begin{tabular}{|r|c|c|c|c|c|c|}
\hline
 $n$ & $N_c^2$ & $N_c^0$ & $N_c^{-2}$ & $N_f N_c$ & $N_f/N_c$ & $N_f^2$ \\
\hline 
$1$ & $4.87(1) \cdot 10^{2}$ & $1.27(4) \cdot 10^{1}$ & $5.63(6) \cdot 10^{-1}$ & $-4.448(9) \cdot 10^{2}$ & $-6.56(2) \cdot 10^{0}$ & $8.34(2) \cdot 10^{1}$ \\
$2$ & $4.539(8) \cdot 10^{1}$ & $-1.38(1) \cdot 10^{0}$ & $-1.91(3) \cdot 10^{-2}$ & $-3.664(4) \cdot 10^{1}$ & $6.67(2) \cdot 10^{-1}$ & $5.511(1) \cdot 10^{0}$ \\
$3$ & $6.60(2) \cdot 10^{0}$ & $-2.83(3) \cdot 10^{-1}$ & $-2.98(7) \cdot 10^{-3}$ & $-5.383(8) \cdot 10^{0}$ & $1.278(5) \cdot 10^{-1}$ & $7.954(2) \cdot 10^{-1}$ \\
$4$ & $1.198(5) \cdot 10^{0}$ & $-5.70(7) \cdot 10^{-2}$ & $-5.6(2) \cdot 10^{-4}$ & $-1.009(2) \cdot 10^{0}$ & $2.54(1) \cdot 10^{-2}$ & $1.5085(6) \cdot 10^{-1}$ \\
$5$ & $2.44(1) \cdot 10^{-1}$ & $-1.21(2) \cdot 10^{-2}$ & $-1.19(6) \cdot 10^{-4}$ & $-2.147(6) \cdot 10^{-1}$ & $5.43(4) \cdot 10^{-3}$ & $3.285(2) \cdot 10^{-2}$ \\
$6$ & $5.34(4) \cdot 10^{-2}$ & $-2.69(6) \cdot 10^{-3}$ & $-2.7(2) \cdot 10^{-5}$ & $-4.94(2) \cdot 10^{-2}$ & $1.22(1) \cdot 10^{-3}$ & $7.777(5) \cdot 10^{-3}$ \\
$7$ & $1.22(1) \cdot 10^{-2}$ & $-6.1(2) \cdot 10^{-4}$ & $-6.4(5) \cdot 10^{-6}$ & $-1.196(5) \cdot 10^{-2}$ & $2.84(3) \cdot 10^{-4}$ & $1.948(1) \cdot 10^{-3}$ \\
$8$ & $2.85(4) \cdot 10^{-3}$ & $-1.42(5) \cdot 10^{-4}$ & $-1.5(2) \cdot 10^{-6}$ & $-3.00(2) \cdot 10^{-3}$ & $6.7(1) \cdot 10^{-5}$ & $5.082(4) \cdot 10^{-4}$ \\
$9$ & $6.8(1) \cdot 10^{-4}$ & $-3.3(2) \cdot 10^{-5}$ & $-3.8(5) \cdot 10^{-7}$ & $-7.75(5) \cdot 10^{-4}$ & $1.62(3) \cdot 10^{-5}$ & $1.367(1) \cdot 10^{-4}$ \\
$10$ & $1.61(4) \cdot 10^{-4}$ & $-7.7(5) \cdot 10^{-6}$ & $-9(2) \cdot 10^{-8}$ & $-2.04(2) \cdot 10^{-4}$ & $3.93(9) \cdot 10^{-6}$ & $3.767(3) \cdot 10^{-5}$ \\
\hline
\end{tabular}
}
\caption{\label{table_moments_y23_colour}
Individual contributions from the different colour factors to the next-to-next-to-leading order
contribution to the $n$-th moment of the three-to-two jet transition distribution.
}
\end{center}
\end{table}

\begin{table}[p]
\begin{center}
{\scriptsize
\begin{tabular}{|r|c|c|c|c|c|c|}
\hline
 $n$ & $N_c^2$ & $N_c^0$ & $N_c^{-2}$ & $N_f N_c$ & $N_f/N_c$ & $N_f^2$ \\
\hline 
$1$ & $2(1) \cdot 10^{4}$ & $-1(1) \cdot 10^{3}$ & $4.0(5) \cdot 10^{2}$ & $5(2) \cdot 10^{3}$ & $-2.3(2) \cdot 10^{3}$ & $3.1(2) \cdot 10^{3}$ \\
$2$ & $1.209(6) \cdot 10^{3}$ & $8.87(6) \cdot 10^{1}$ & $6.81(4) \cdot 10^{0}$ & $-1.088(2) \cdot 10^{3}$ & $-6.57(1) \cdot 10^{1}$ & $2.665(1) \cdot 10^{2}$ \\
$3$ & $3.816(4) \cdot 10^{2}$ & $2.66(5) \cdot 10^{0}$ & $-2.91(9) \cdot 10^{-2}$ & $-3.076(2) \cdot 10^{2}$ & $-3.73(8) \cdot 10^{-1}$ & $5.2590(4) \cdot 10^{1}$ \\
$4$ & $1.234(1) \cdot 10^{2}$ & $-2.48(2) \cdot 10^{0}$ & $-8.44(4) \cdot 10^{-2}$ & $-9.574(6) \cdot 10^{1}$ & $1.423(4) \cdot 10^{0}$ & $1.5030(2) \cdot 10^{1}$ \\
$5$ & $4.274(7) \cdot 10^{1}$ & $-1.416(9) \cdot 10^{0}$ & $-3.22(2) \cdot 10^{-2}$ & $-3.362(3) \cdot 10^{1}$ & $7.21(2) \cdot 10^{-1}$ & $5.2026(7) \cdot 10^{0}$ \\
$6$ & $1.574(3) \cdot 10^{1}$ & $-6.32(4) \cdot 10^{-1}$ & $-1.20(1) \cdot 10^{-2}$ & $-1.290(1) \cdot 10^{1}$ & $3.167(9) \cdot 10^{-1}$ & $2.0142(4) \cdot 10^{0}$ \\
$7$ & $6.10(2) \cdot 10^{0}$ & $-2.68(2) \cdot 10^{-1}$ & $-4.83(5) \cdot 10^{-3}$ & $-5.281(7) \cdot 10^{0}$ & $1.380(5) \cdot 10^{-1}$ & $8.377(2) \cdot 10^{-1}$ \\
$8$ & $2.458(9) \cdot 10^{0}$ & $-1.13(1) \cdot 10^{-1}$ & $-2.09(3) \cdot 10^{-3}$ & $-2.269(4) \cdot 10^{0}$ & $6.12(2) \cdot 10^{-2}$ & $3.6598(9) \cdot 10^{-1}$ \\
$9$ & $1.025(5) \cdot 10^{0}$ & $-4.78(6) \cdot 10^{-2}$ & $-9.6(2) \cdot 10^{-4}$ & $-1.013(2) \cdot 10^{0}$ & $2.79(1) \cdot 10^{-2}$ & $1.6566(5) \cdot 10^{-1}$ \\
$10$ & $4.40(2) \cdot 10^{-1}$ & $-2.04(3) \cdot 10^{-2}$ & $-4.67(9) \cdot 10^{-4}$ & $-4.66(1) \cdot 10^{-1}$ & $1.299(7) \cdot 10^{-2}$ & $7.700(3) \cdot 10^{-2}$ \\
\hline
\end{tabular}
}
\caption{\label{table_moments_sqrtheavyjetmass_colour}
Individual contributions from the different colour factors to the next-to-next-to-leading order
contribution to the $n$-th moment of the square root of the heavy jet mass distribution.
}
\end{center}
\end{table}

\clearpage

\begin{table}[p]
\begin{center}
{\scriptsize
\begin{tabular}{|r|c|c|c|c|}
\hline
 $n$ & $\kappa=10^{-3}$ & $\kappa=10^{-5}$ & $\kappa=10^{-7}$ & $\kappa=10^{-9}$ \\
\hline 
 $1$ & $ -2.8 \cdot 10^{1}$ & $1.4 \cdot 10^{ 1}$ & $1.2 \cdot 10^{ 0}$ & $4.9 \cdot 10^{-2}$ \\
 $2$ & $ -3.0 \cdot 10^{-2}$ & $4.5 \cdot 10^{-5}$ & $4.7 \cdot 10^{-8}$ & $2.1 \cdot 10^{-11}$ \\
 $3$ & $ -2.0 \cdot 10^{-5}$ & $2.6 \cdot 10^{-10}$ & $2.9 \cdot 10^{-15}$ & $1.3 \cdot 10^{-20}$ \\
 $4$ & $ -1.5 \cdot 10^{-8}$ & $1.8 \cdot 10^{-15}$ & $2.1 \cdot 10^{-22}$ & $9.7 \cdot 10^{-30}$ \\
 $5$ & $ -1.2 \cdot 10^{-11}$ & $1.4 \cdot 10^{-20}$ & $1.6 \cdot 10^{-29}$ & $7.6 \cdot 10^{-39}$ \\
 $6$ & $ -1.0 \cdot 10^{-14}$ & $1.1 \cdot 10^{-25}$ & $1.3 \cdot 10^{-36}$ & $6.3 \cdot 10^{-48}$ \\
 $7$ & $ -8.6 \cdot 10^{-18}$ & $9.4 \cdot 10^{-31}$ & $1.1 \cdot 10^{-43}$ & $5.4 \cdot 10^{-57}$ \\
 $8$ & $ -7.5 \cdot 10^{-21}$ & $8.1 \cdot 10^{-36}$ & $9.9 \cdot 10^{-51}$ & $4.7 \cdot 10^{-66}$ \\
 $9$ & $ -6.6 \cdot 10^{-24}$ & $7.1 \cdot 10^{-41}$ & $8.8 \cdot 10^{-58}$ & $4.1 \cdot 10^{-75}$ \\
$10$ & $ -5.9 \cdot 10^{-27}$ & $6.4 \cdot 10^{-46}$ & $7.9 \cdot 10^{-65}$ & $3.7 \cdot 10^{-84}$ \\
\hline
\end{tabular}
}
\caption{\label{table_log_terms}
Contribution to the next-to-next-to-leading order coefficient
$C_{(1-T),n}$ for the $n$-th moment of the thrust distribution obtained from integrating the logarithmic terms
from $0$ to $\kappa$ for various values of $\kappa$.
}
\end{center}
\end{table}

\end{document}